\documentstyle[eqsecnum,aps,psfig,twocolumn]{revtex} 

\begin{document}
\draft
\preprint{ }
\title{Cosmological constraints from lensing statistics and supernovae on the cosmic equation of state}
\author{Ioav Waga$^1$ and Ana Paula M. R. Miceli$^{1,2}$}
\address{$^1$Universidade Federal do Rio de Janeiro, Instituto de F\'\i sica, C. P. 68528, 21945-970, Brazil.\\
$^2$NASA/Fermilab Astrophysics Center, Fermi National Accelerator Laboratory 
PO Box 500, Batavia IL 60510 - USA
}
\date{\today}
\maketitle
\begin{abstract}
We investigate observational constraints from lensing statistics and high-z type Ia supernovae on flat cosmological models with nonrelativistic matter and an exotic fluid with equation of state, $p_x=(m/3 -1)\rho_x$. We show that agreement with both tests at the $68$\% confidence level is possible if the parameter $m$ is low ($m \lesssim 0.85$) and $0.24 \lesssim \Omega_{m0} \lesssim 0.38$ with lower values of $\Omega_{m0}$ corresponding to higher $m$. We find that a conventional cosmological constant model with $\Omega_{m0}\simeq 0.33$ is the best fit model of the combined likelihood.
\end{abstract}
\pacs{PACS number(s): 98.80.Hw}

\narrowtext

\section{Introduction}
\label{sec:level1}
Flat cosmological models with a cosmological constant are currently the favorite candidates to describe the dynamics of the universe. These models are theoretically appealing  because $\Lambda$ helps to reconcile
inflation with dynamic estimates for the matter density parameter ($\Omega_{m0}$). Further, they fit nicely some observational data, as for instance those coming from the classical magnitude-redshift test, where SneIa are used as standard candles, and also from the favored location of the first acoustic peak of the cosmic microwave background radiation (CMBR) angular power spectra.\cite{rie98,han98} 

Historically, Einstein himself was the first one to include a $\Lambda$-term in the general relativity field equations in order to make them compatible with a static universe. Several times $\Lambda$ was introduced in cosmology but later on discarded when improved data became available. The preliminary results from SneIa almost excluded $\Lambda$, so it is curious to note that this time better data is supporting $\Lambda$ instead of rejecting it \cite{per97,per98}. 
From the aesthetic point of view, if compared, for instance, to the Einstein-de Sitter models, $\Lambda$-models seem ugly. A new parameter is introduced in the theory and in order to dominate the dynamics of the universe only in recent times, this parameter should have a very small value ($\Lambda \lesssim 10^{-56}\;cm^{-2}$), which is 50 to 120 orders of magnitude below the estimate given by quantum field theory. In fact, in part to alleviate this problem dynamical-$\Lambda$ models  were originally suggested. 
There are also observational motivations for considering dynamical $\Lambda$ models instead of the constant one. 
For instance, in these models the COBE-normalized amplitude of the mass power spectrum is in general lower than in the conventional constant-$\Lambda$ model, in accordance with observations \cite{cob97}. Further, the distance to an object with redshift $z$ is smaller than the distance to the same object in a constant-$\Lambda$ model (assuming the same value of $\Omega_{m0}$). So, constraints coming from lensing statistics are weaker in these models \cite{rat92,blo96}.

The dynamical-$\Lambda$ models present in the literature can schematically be divided in three types: scalar field \cite{sca,jos95,cob97,cal98,via98,jos98}, x-fluid \cite{xflu,tur97,chi97,hu98,whi98} and decaying-$\Lambda$ laws \cite{lamb,sil94,blo96,sil97}. 
A phenomenological decaying-$\Lambda$ law model in which $\Lambda$ decreases as
$\Lambda \propto a^{-m}$ [here $a$ is the scale factor of the Friedman-Robertson-Walker (FRW) metric and $m$ is a constant ($0\leq m <3$)] was suggested in Refs.\cite{sil94,blo96,sil97}. It was observed that the Einstein equations for these models are the same if instead of a $\Lambda$-term, it would be considered (beside matter and radiation) a x-fluid with equation of state, $p_x=\left( \frac m3-1\right) \rho _x$. In spite of the similarity at the level of Einstein equations, these two phenomenological models are different. For instance, in the case of a decaying $\Lambda$-term, matter is created as a result of the decaying vacuum, while in the exotic fluid description the x-component is conserved. In this sense the last approach is more conventional and closer to the description of dynamical-$\Lambda$ in terms of a scalar field rolling down a potential without coupling to other fields. 

In previous work \cite{sil94,sil97}, as a first approximation, it was assumed that the x-component is smoothly distributed. We were motivated by the fact that if the x-component clumps on scales $\sim 10-20$ $h^{-1}$Mpc it would be detected by dynamical measurements and this is not observed. Another concern is that if the perturbed pressure is negative (as is the background one), the fluid sound velocity would be imaginary, small scales would grow exponentially and the system would be highly unstable \cite{fab97,tur97,cal98}. However, as observed in Refs. \cite{chi97,cal98,hu98}, the fact that the background pressure is negative not necessarily implies imaginary sound speed. Furthermore, Caldwell {\it et al.} \cite{cal98} (see also Ref.\cite{via98}) pointed out that the smoothness assumption is gauge dependent. However, if for instance $\Lambda$ is modeled by a scalar field, perturbations will only give an appreciable effect at large scales. For scales well inside the horizon, smoothness is a good approximation.\cite{whi98}.  

In this paper we shall deal with two cosmological tests:  gravitational lensing statistics and the SneIa magnitude redshift test. The strongest observational support for an accelerated universe comes from SneIa. This test can be considered the main motivation for introducing some kind of exotic matter with negative pressure. The cosmological constant is the simplest possibility, but not the unique. On the other hand, most lensing statistics analysis give lower values for $\Lambda$ and we find interesting to compare the predictions of these two important tests. Here we consider the special case where the exotic component is a x-fluid with constant equation of state and that is smooth on scales smaller than horizon. We show that agreement with both tests at the $68$\% confidence level is possible if the parameter $m$ is low ($m \lesssim 0.85$) and $0.24 \lesssim \Omega_{m0} \lesssim 0.38$ with lower values of $\Omega_{m0}$ corresponding to higher $m$. The best fit model of the combined likelihood has $m=0$ (cosmological constant) and $\Omega_{m0}\simeq 0.33$ ($\Omega_{\Lambda0}\simeq 0.67$).

This paper is organized as follows. 
In Sec. II the basic field equations and distance formulae are presented. In Sec. III we obtain constraints on the models from lensing statistics. Constraints from high redshift SneIa  are obtained in Sec. IV. We also present a combined likelihood analysis of both tests in this section. In Sec. V our main conclusions are stressed out.

\section{Field equations and distance formulae}

In this paper we consider spatially flat, homogeneous, and 
isotropic cosmologies
with nonrelativistic matter and an exotic x-fluid with 
equation of state, $p_x=\left( \frac m3-1\right) \rho _x$ (or 
equivalently a time-dependent $\Lambda$ term such that
$\Lambda \propto a^{-m}$). Since we are mostly interested on effects that occurred at redshift $z<5$ we neglect radiation. We consider that nonrelativistic matter and x-fluid  are separately conserved ($\rho_m \propto a^{-3}$ and $\rho_x \propto a^{-m}$).

The Einstein equations for the models we are considering are:
\begin{equation}
\left( \frac{\stackrel{.}{a}}a\right) ^2
=\Omega _{m0}H_0^2\left( \frac{a_0}a\right) ^3 +
\Omega_{x0}H_0^2
\left( \frac{a_0}a\right) ^m 
\end{equation}
\noindent and 
\begin{equation}
\frac{\stackrel{..}{a}}a=-\frac 12\Omega _{m0}H_0^2
\left( \frac{a_0}a\right)^3 +
\frac{(2-m)}2\Omega_{x0}H_0^2\left( \frac{a_0}a\right) ^m, \end{equation}
\noindent
where $\Omega _{m0} = 1 - \Omega _{x0} 
$ is the matter density parameter and $H_0$ is the present value of
the Hubble parameter.

In the next sections we shall use two concepts of cosmological distances, the angular diameter distance and the luminosity distance. So, we end this section presenting their definitions and showing some expressions we use in our computations.
 
Consider that photons are emitted by a source with coordinate $r=r_{1}$  at
time $t_{1}$ and are received at time $t_{0}$ by an observer located at coordinate $r=0.$
The emitted radiation will follow null 
radial geodesics on which $\theta $ and $\phi $  are
constant. The comoving distance of the source is defined by:
\begin{equation}
\chi = \int^{t_{0}}_{t_{1}} \frac{c dt}{a(t)} \;,
\end{equation} 
that in flat space is equal to $r_1$.
The present value of the scale factor times the comoving 
distance,  gives  the  proper
distance, $d(0,z)$, between the source and the observer, 
\begin{equation}
d(0,z) = a_0 \int^{a_0}_{a(t)} \frac{c da}{\dot a a},
\end{equation}
and in our case it reduces to the following expression, 
\begin{equation}
d(0,z)=c{H_{0}}^{-1}\int_{0}^{z}\frac{dy}{\sqrt{\Omega _{m0}(1+y)^{3}
+ (1-\Omega_{m0}) (
1+y )^{m}}} .  \nonumber \\ 
\end{equation}    

The  luminosity distance of a light source is defined in such a way as to generalize to an
expanding and curved space the inverse-square law of  brightness  valid  in  a static Euclidean space,
\begin{equation}
d_{L}(0,z)= \left(\frac{\cal{L}}{4\pi \cal{F}}\right)^{1/2}= r_{1} a_0 (1 + z).
\end{equation}
\noindent 
Here $\cal{L}$ is the total energy emitted by  the  source
per  unit  time in its rest frame (absolute luminosity), $\cal{F}$ is the measured flux (energy per unit time per unit area measured by a detector). For flat models, 
$d(0,z) = a_0 r_{1}$ and $d_{L}$ can therefore be written as
\begin{equation}
d_{L}(0,z) = (1 + z) d(0,z).
\end{equation}
\noindent 

The angular diameter of  a  light  source  of  proper  diameter $D$ (here
supposed to be redshift independent) at $r=r_{1}$ and $t=t_{1}$, 
observed at $r=0$ and $t=t_{0}$
is given by
\begin{equation}
\delta  = \frac{D}{a(t_{1})r_{1}} = \frac{D(1+z)}{d(0,z)} 
= \frac{D(1+z)^2}{d_{L}(0,z)} .
\end{equation}
\par
The angular diameter distance is defined  as  the  ratio  of  the  source
diameter to its angular diameter, 
$i.e.$, it  is  the  distance  that  would  be
attributed to the light source if it were in a Euclidean space,
\begin{equation}
d_{A}(0,z) = \frac{D}{\delta } =  d(0,z) (1+z)^{-1} 
= d_{L}(0,z) (1+z)^{-2} .
\end{equation}
It is convenient to write the angular diameter distance for our models, between the redshift $z_L$ and $z_S$,  
\begin{eqnarray}
d_A(z_L,&&z_S) =\nonumber \\&& \frac{c{H_{0}}^{-1}}{1+z_S}\int_{z_L}^{z_S}\frac{dy}{\sqrt{\Omega _{m0}(1+y)^{3}
+ (1-\Omega_{m0}) (
1+y )^{m}}} .\nonumber \\\end{eqnarray}

\narrowtext
\section{Constraints from lensing statistics}

We start defining the following likelihood function, \cite{koch93}
\begin{equation}
{\mathcal L}_{lens}=\prod_{i=1}^{N_U} (1-p^{'}_{i}) \prod_{j=1}^{N_L} p^{'}_{j} \prod_{k=1}^{N_L} p^{'}_{ck}.
\end{equation}
Here $N_L$ is the number of quasars that have multiple image, $N_U$ is the number of quasars that don't have, $p^{'}_i\ll1$ is the probability that quasar $i$ is lensed and $p^{'}_{ck}$ is the configuration probability, that we shall consider as the probability that  quasar $k$ is lensed with the observed image separation. To perform the statistical analysis we use data from
the HST Snapshot survey (498 high luminous quasars (HLQ), the Crampton survey (43 HLQ), the Yee survey (37 HLQ), the ESO/Liege survey (61 HLQ), The HST GO observations (17 HLQ), the CFA survey (102 HLQ) , and
the NOT survey (104 HLQ) \cite{mao}. We considered a total of 862 ($z > 1$) high luminous optical quasars plus 5 lenses.

The differential probability, $d\tau$, that a line of sight intersects a galaxy at redshift $z_L$ in the interval $dz_L$ from a population with number density $n_G$ is,
\begin{equation}
d\tau = \frac{\mbox{differential light travel distance}}{\mbox{mean free path}} = \frac{c dt}{1/n_G \pi a_{cr}^2}
\end{equation}
where $a_{cr}$ is the maximum distance of the lens from the optical axes for which multiple images are possible. It is a function of the angular diameter distance between observer and lens, lens and source, observer and source and it also depends on the lens model.

We use a singular isothermal sphere (SIS) as the lens model. Hence,
$$
a_{cr}=\frac{d_A(0,z_L)d_A(z_L,z_S)}{d_A(0,z_S)} \alpha_0,
$$
and the bending angle is $\alpha_0 = 4\pi$($\frac{\sigma_{||}}{c})^2 \simeq 1.5^{''}(\frac{\sigma_{||}}{225\mbox{Km/s}})^2$ with $\sigma_{||}$ standing for the one component velocity dispersion. We assume conserved comoving number density of lenses, $n_G=n_0(1+z)^3$ and a Schechter \cite{sch76} form for the galaxy population. So,
\begin{equation}
n_0 = \int^{\infty}_{0} n_{*}\left(\frac{L}{L^{*}}\right)^{\alpha}
\exp \left(-\frac{L}{L^{*}}\right)
\frac{dL}{L^{*}}.
\end{equation}
and we take $n_{*}=1.4 \pm 0.17\;h^310^{-2}\mbox{Mpc}^{-3}$ and $\alpha=-1.0 \pm 0.15$ \cite{lov94}. The division of the luminosity function by galaxy type is taken from Marzke {\it et al.} \cite{mar94}, $n_{e}=0.61 \pm 0.21\;h^310^{-2}\mbox{Mpc}^{-3}$ for early type galaxies and $n_{s}=0.79 \pm 0.26\; h^3 10^{-2} \mbox{Mpc}^{-3}$ for spirals \cite{koch96}. As a conservative approach biased pro higher values of $\Omega_{\Lambda}$, we do not consider lensing by spiral galaxies. We assume that the luminosity satisfies the Faber-Jackson relation \cite{fab76}, ${L}/{L^{*}}=({\sigma_{||}}/{\sigma_{||}^{*}})^{\gamma}$, with 
$\gamma=4$ and take $\sigma_{||}^{*}= 225$ Km/s.

The total optical depth ($\tau$), obtained by integrating $d\tau$ along the line of sight from $0$ to $z_S$, can be expressed analytically,
\begin{equation}
\tau(z_S)=\frac{F}{30}\big(d_A(0,z_S) (1+z_S)\big)^3 (c H_0^{-1})^{-3},
\end{equation}
where $F=16\pi^3 n_e (c H_0^{-1})^3 (\sigma_{||}^{*}/c)^4 \Gamma (1+\alpha+4/\gamma)\simeq0.026$ measures the effectiveness of the lens in producing multiple images \cite{tog}.

It is important to include two corrections to the optical depth: magnification bias and selection function due to finite resolution and dynamic range \cite{koch93}. 

Since lensing increase the apparent brightness of a quasar and since there are more faint quasars than bright ones, there will be over representation of lensed quasars in a flux limited sample. The bias factor is given by \cite{ft91,koch93,koch96}
\begin{equation}
{\mathbf B}(m,z)=M_0^2\;{\mathcal B}(m,z,M_0,M_2)
\end{equation}
where
\begin{eqnarray}
&&{\mathcal B}(m,z,M_1,M_2)= \nonumber \\&&2 \left(\frac{dN_q}{dm}\right)^{-1} \int^{M_2}_{M_1}\frac{dM}{M^3}\;\frac{dN_q}{dm}(m+2.5\log{M},z).
\end{eqnarray}
Since we are modeling the lens by a SIS profile, $M_0=2$, and we use $M_2=10^4$ in the numerical computation.

We use the following expression for the quasar luminosity function \cite{koch96}
\begin{equation}
\frac{dN_q}{dm} \propto \left( 10^{-a(m-\overline{m})} + 10^{-b(m-\overline{m})}\right)^{-1},
\end{equation}
where
\begin{eqnarray}
\overline{m} = \left\{ 
\begin{array}{lll}
m_0 + (z+1) & \;\;{\rm for}\;\;\;\;\;\;\;\;\;\;\; z<1, &  \\ 
m_0 & \;\;{\rm for}\;\;\;\; 1<z<3, &  \\ 
m_0 -0.7 (z-3) & \;\;{\rm for}\;\; \;\;\;\;\;\;\;\;\; z>3, & 
\end{array}
\right.
\end{eqnarray}
and we assume $a=1.07$, $b=0.27$ and $m_0=18.92$. The magnification corrected probabilities are
\begin{equation}
p_i = \tau(z_i) {\mathbf B}(m_i,z_i)
\end{equation}

Finally we have to consider the selection function due to finite resolution and dynamic range. It can be shown that the selection function corrected probabilities are: \cite{koch93}
\begin{equation}
p^{'}_{i}(m,z) = p_i \frac{\int d\theta p_c(\theta){\cal B}(m,z,M_f(\theta), M_2)}{{\cal B}(m,z,M_0, M_2)}
\end{equation}
and
\begin{equation}
p^{'}_{ci} = p_{ci}(\theta) \left( \frac{p_{i}}{p^{'}_{i}} \right) \frac{{\cal B}(m,z,M_f(\theta),M_2)} {{\cal B}(m,z,M_0, M_2)},
\end{equation}
where \cite{pc}

\begin{eqnarray}
p_c(\theta) &=&  \frac{F}{\tau(z_S)}\int^{z_S}_0 (1+z_L)^3\left(\frac{d_A(0,z_S)d_A(z_L,z_S)}{cH{_0}^{-1}d_A(0,z_S)}\right)^2
\nonumber \\ &\times&  8 \pi \left(\frac{\sigma_{||}^{\star}}{c}\right)^2 
\left(-\frac 1{cH_{0}^{-1}}\frac{cdt}{dz_L}\right) \nonumber \\
&\times&
\frac{\gamma/2}{\Gamma(\alpha+1+\frac4\gamma)} 
\left(\frac{d_A(0,z_S)}{d_A(z_L,z_S)}\frac{\theta }{ 8 \pi \left(\frac{\sigma_{||}^{\star}}{c}\right)^2}
\right)^{\frac{\gamma}{2}(\alpha+1+\frac4\gamma)}\nonumber \\ &\times&
\exp{\left[-\left(\frac{d_A(0,z_S)}{d_A(z_L,z_S)}\frac{\theta }{ 8 \pi \left(\frac{\sigma_{||}^{\star}}{c}\right)^2}\right)^{\frac{\gamma}{2}} \right]}\frac{1}{\theta} dz_L,
\end{eqnarray} 
\begin{equation}
M_f(\theta)= M_0 \frac{1+f}{f-1}, \;\;\;\; f>1
\end{equation}
and
\begin{equation}
f=f(\theta)=10^{0.4 \Delta m(\theta)}.
\end{equation}
To simplify computation we use two selection functions, one for the HST observations and another one for all the ground based surveys \cite{sf}. Using more accurate selection functions for each ground based observations separately have little statistical effect. 

Recently Falco {\it et al.} \cite{fal98} observed that statistical lensing analysis based on optical and radio observations can be reconciled if the existence of dust in E/SO galaxies is considered. In our computation we assume a mean extinction of $\Delta m$ =$0.5$ mag as suggested by their estimates.

By expressing ${\cal{L}}_{lens}$ as a function of the parameters $m$ and $\Omega_{m0}$ we obtained the maximum of the likelihood function (${\cal{L}}_{lens}^{max}$) and formed the ratio $l={\cal{L}}_{lens}/{\cal{L}}_{lens}^{max}$. It can be shown that with two parameters, the distribution of $-2\ln{l}$ tends to a $\chi ^2$ distribution with two degrees of freedom \cite{ken77,koch93}. In two dimensions the 68\%($1 \sigma$) and 95.4\% ($2\sigma$) confidence levels are where the likelihood is $60.7\%$ and $13.5\%$, respectively, of the peak likelihood.

In Fig.$1$ we plot contours of constant likelihood ($95.4\%$ and $68\%$) in the two parameter space ($m$, $\Omega_{m0}$). The maximum of the likelihood occurs for $m\simeq 2.4$ and $\Omega_{m0}= 0$. The same approach when applied to constant $\Lambda$ models (since $m=0$ we now have only one degree of freedom) gives: $\Omega_{\Lambda} \lesssim 0.76$ (or $\Omega_{m0} \gtrsim 0.24$) at $2\sigma$, $1 \gtrsim \Omega_{m0} \gtrsim 0.39$ at $1\sigma$ with a
best fit at $\Omega_{m0} \simeq 0.62$. For comparison, without considering extinction, in the case $m=0$, we get:
$\Omega_{\Lambda} \lesssim 0.55$ at $2\sigma$, that is slightly more conservative than Kochanek's \cite{koch93} $\Omega_{\Lambda} \lesssim 0.66$ at $2\sigma$. 

In Fig.$2$ contours of constant number of multiple images as a function of the parameters $m$ and $\Omega_{m0}$ are displayed for the case in which $\Delta m =0.5$ mag extinction is considered. Comparing the two figures we see that the $1\sigma$ contour of Fig. $1$ corresponds roughly to the contour in Fig. $2$ where $9$ multiple images are expected. For the best fit model in two dimensions the predicted number of multiple images is $\simeq 5.3$. 

The results obtained in this section are more accurate than those presented in Ref.\cite{sil97}. We now took into account magnification bias in the configuration probability and considered the selection function due to finite resolution and dynamic range. In the present approach extinction is also considered, so the constraints are less restrictive with respect to $\Lambda$ than those obtained in Ref.\cite{sil97}  .

\begin{figure} \hspace*{0.1in}
\psfig{file=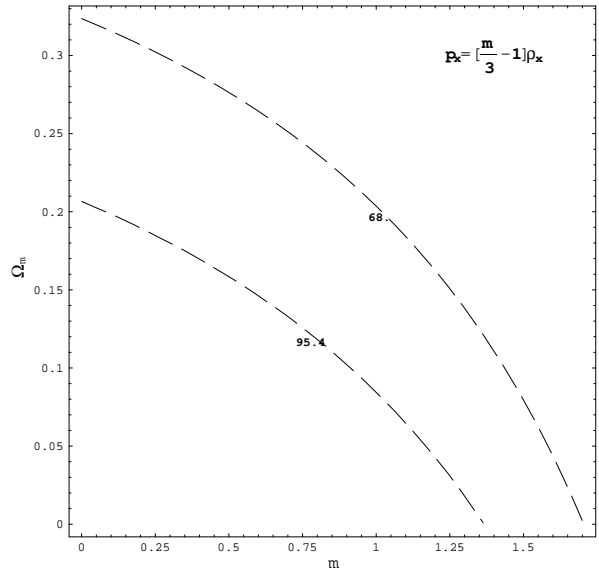,height=7.5cm,width=7.8cm}
\vspace*{0.4in}
\caption{Contours of constant likelihood ($95.4\%$ and $68\%$) arising from  lensing statistics.}
\end{figure}

\begin{figure} \hspace*{0.1in}
\psfig{file=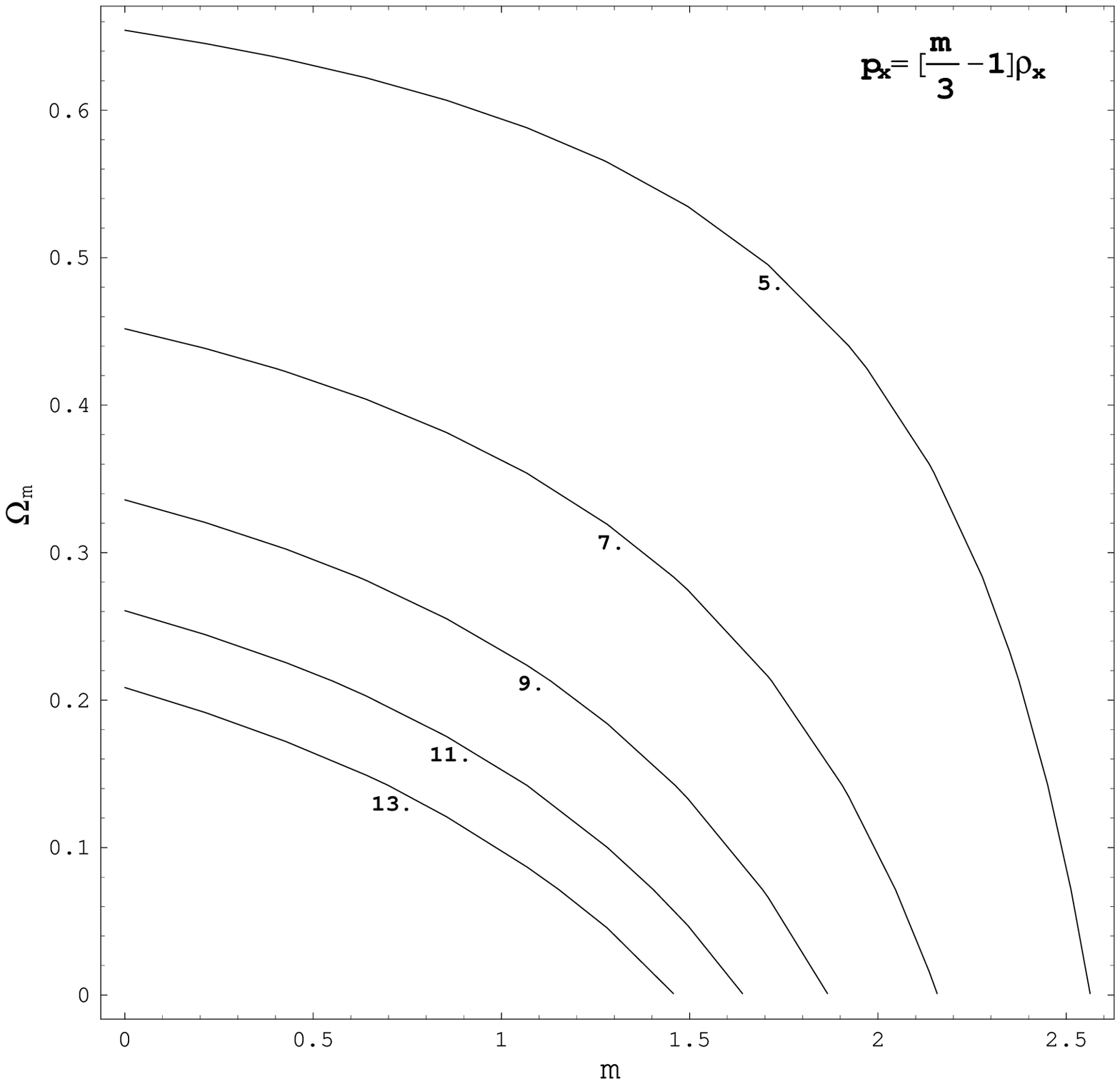,height=7.5cm,width=7.8cm}
\vspace*{0.4in}
\caption{Contours of constant number of multiple images as a function of the parameter space.}
\end{figure}

\narrowtext
\section{Constraints from high-redshift type Ia Supernovae}
There are two major ongoing programs to systematically 
search and study high-z supernovae. Although the very preliminary results indicated a low value for the cosmological constant ( $\Omega_{\Lambda} < 0.51$ at the $95\%$ confidence level) \cite{per97}, more recent analysis with larger sample of supernovae, now points to a different direction. Now the data indicate an accelerated expansion such that $\Omega_{\Lambda} \sim 0.7$, $\Omega_{m0} \sim 0.3$ and strongly supports a flat Universe.\cite{rie98,per98}.

In this section we update the constraints on the equation of state obtained by Silveira \& Waga \cite{sil97}, where the first data from the Supernovae Cosmology Project was used \cite{per97}. We now consider data from the High-z Supernovae Search Team. We use the 27 low-z and 10 high-z SneIa (we include SN97ck) reported in Riess {\it et al.} \cite{rie98} and consider data with the MLCS \cite{rie96,rie98} method applied to the supernovae light curves. The results for spatially flat models that we present here, are similar to those obtained by Garnavich {\it et al.} \cite{gar98}.

Following a procedure similar to that described in Riess\,{\it et al.}\cite{rie98}, we determine the cosmological parameters $m$ and $\Omega_{m0}$ through a $\chi ^2$ minimization neglecting the unphysical region $\Omega_{m0}<0$. To simplify computation we fix the Hubble parameter to $H_0 = 65.2$ km/s Mpc$^{-1}$ \cite{rie98}, but the results are independent of this choice for $H_0$ \cite{rie98,gar98}. We use
\begin{equation}
\chi^{2}_{sne}(\Omega_{m0},m) = \sum_{i=1}^{37} \frac{\left(\mu_p(z_i,\Omega_{m0},m) - \mu_{0,i}\right)^2}{\sigma_{\mu_{0,i}}^2+\sigma_{vi}^2},
\end{equation}
where
\begin{equation}
\mu_p = 5 \log{d_L} + 25, 
\end{equation}
is the distance modulus predicted by each model, $\mu_0$ is the observed (after corrections) distance modulus, $\sigma_{\mu_0}$ its uncertainty and $\sigma_v$ is the dispersion in galaxy redshift due to peculiar velocities. Following \cite{rie98} we use $\sigma_{vi}=\frac{5}{\ln{10}}\frac{200km/s}{cz_i}$ and for high-z SneIa with z not derived from emission lines in the host galaxy, we add $2500$ km/s in quadrature to $200$ km/s (see Table $1$ in \cite{rie98}).

\begin{figure} \hspace*{0.1in}
\psfig{file=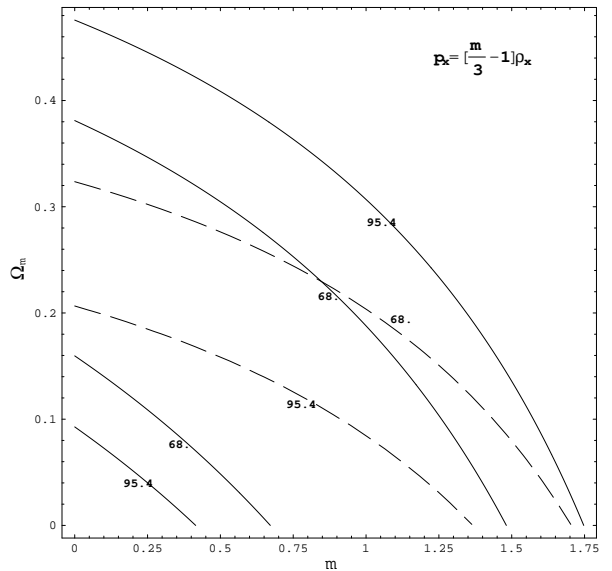,height=7.5cm,width=7.8cm}
\vspace*{0.4in}
\caption{Contours of constant likelihood ($95.4\%$ and $68\%$) arising from lensing statistics (dashed lines) and type Ia supernovae are shown.}
\end{figure}

In Fig.$3$ contours of constant likelihood 95.4\% ($2\sigma$) and 68\% ($1\sigma$) arising from the $\Delta\chi^2_{sne}$ analysis are displayed together with those from lensing (dashed lines). For SneIa the peak of the likelihood is located at $m\simeq 1.1$ and $\Omega_{m0}=0$. If we fix $m=0$ we get $\Omega_{m0} = 0.25 \pm 0.08$ ($1\sigma$) (for comparison Riess {\it et al.} obtained $\Omega_{m0} = 0.24 \pm 0.1$). From the figure it is clear that there is a region in the parameter space (the region inside the triangle with vertices ($m\simeq0.85$,$\Omega_{m0}\simeq 0.24$), ($m= 0$,$\Omega_{m0}\simeq 0.32$) and ($m=0$,$\Omega_{m0}\simeq 0.38$)) such that all points are inside the $1\sigma$ ($68\%$) confidence region of both tests.  

In Fig. $4$ we display contours ($95.4\%$ and $68\%$) of the combined (lensing plus SneIa) likelihood. For the combined $\chi^2$ analysis we used $\chi^2_{tot}= \Delta\chi^2_{sne} - 2\ln{l}$, with $l={\cal{L}}_{lens}/{\cal{L}}_{lens}^{max}$ as defined in Sec. III. Although the peak of the likelihood for each test separately occurs at $\Omega_{m0}=0$, the maximum of the combined likelihood occurs at $m=0$ (cosmological constant) and $\Omega_{m0}\simeq 0.33$. Note that best fit models of the combined likelihood are in accelerated expansion ($q_0<0$). Models with $m=2$ (cosmic strings \cite{xflu}) and any value of $\Omega_{m0}$ are at more than $99\%$ c.l. away from the peak of the likelihood. We observe that if, for instance, we take $h=0.65$ and $\Omega_Bh^2=0.02$, the CMBR first acoustic peak ($\ell_{\rm peak}$), for models with $m$ and $\Omega_{m0}$ inside the $1\sigma$ allowed region in Fig $4$, will have $\ell_{\rm peak}$ values between $\simeq 215$ and $230$, (see Fig. $4$ in Ref.\cite{whi98}), that are close to the current best values for $\ell_{\rm peak}$ obtained from CMBR data \cite{han98}. Models with parameters $m$ and $\Omega_{m0}$ in this region are in agreement with the current CMBR data as well. Constraints from observations of clusters  suggest $\Omega_{m0} > 0.14$ \cite{car97}. In Fig. $4$ we display this constrain as a dotted line. Models above this line are preferred.

\begin{figure} \hspace*{0.1in}
\psfig{file=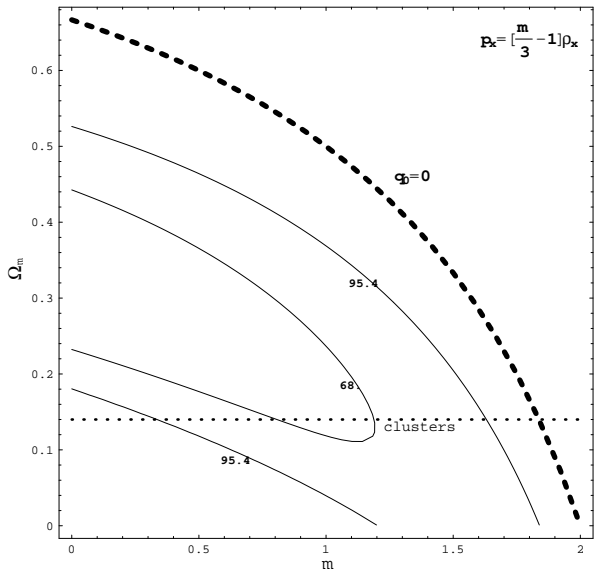,height=7.5cm,width=7.8cm}
\vspace*{0.4in}
\caption{Contours of combined likelihood ($95.4\%$ and $68\%$) arising from  lensing statistics and type Ia supernovae are shown.}
\end{figure}

\narrowtext
\section{Summary}

We have studied observational constraints from lensing statistics and high-z SneIa on spatially flat cosmological models whose matter content is nonrelativistic matter plus an exotic fluid with equation of state, $p_x=(\frac m3 -1)\rho_x$. 
Using a lensing approach, where extinction is considered, and that takes into account magnification bias and the selection function due to finite resolution and dynamic range in the configuration probability, we obtained the $68\%$ and $95.4\%$ confidence level on the parameters $m$ and $\Omega_{m0}$. The present results update those obtained by Silveira \& Waga \cite{sil97}. We also update the SneIa constraints on the equation of state obtained in Ref. \cite{sil97}, where the first data from the Supernovae Cosmology Project was used. We considered data from the High-z Supernovae Search Team, the 27 low-z and 10 high-z SneIa reported in Ref. \cite{rie98}. We used data with the MLCS method applied to the supernovae light curves. We showed that agreement with both tests at the $68$\% confidence level is possible if the parameter $m$ is low ($m \lesssim 0.85$) and $0.24 \lesssim \Omega_{m0} \lesssim 0.38$ with lower values of $\Omega_{m0}$ corresponding to higher $m$. We observed that best fit models of the combined likelihood are in accelerated expansion ($q_0<0$). We obtained that a conventional cosmological constant model with $\Omega_{m0}\simeq 0.33$ ($\Omega_{\Lambda0}=0.67$) is the best fit model of the combined likelihood.

\acknowledgments
We thank Adam Riess, Albert Stebbins, Chris Kochanek, Josh Frieman, Scott Dodelson and Varun Sahni for useful conversations.
This work was supported in part by the Brazilian agencies CAPES, CNPq and FAPERJ.

\end{document}